%% file: bare_jrnl.tex
\documentclass[journal,twocolumn]{IEEEtran}
\makeatletter
\def\endthebibliography{%
  \def\@noitemerr{\@latex@warning{Empty `thebibliography' environment}}%
  \endlist
}
\usepackage{authblk}

\input{package}
\input{lstingpackage}

\begin{document}

\title{Feature Location Benchmark for Decomposing and Reusing Android Apps}

\author[1]{Yutian Tang}
\author[1]{Hao Zhou}
\author[2]{Zhou Xu}
\author[1]{Xiapu Luo}
\author[3]{Yan Cai}
\author[4]{Tao Zhang}
\affil[1]{Department of Computing, The Hong Kong Polytechnic University}
\affil[2]{School of Big Data and Software Engineering, Chongqing University}
\affil[3]{State Key Laboratory of Computer Science, Institute of Software, Chinese Academy of Sciences}
\affil[4]{Faculty of Information Technology, Macau University of Science and Technology}
\affil[*]{Corresponding Author: csxluo@comp.polyu.edu.hk}

\markboth{Journal of \LaTeX\ Class Files}%
{\MakeLowercase{\textit{Tang et al.}}: Feature Location Benchmark for Decomposing and Reusing Android Apps}
\maketitle

\begin{abstract}
Software reuse enables developers to reuse architecture, programs and other software artifacts. Realizing a systematical reuse in software brings a large amount of benefits for stakeholders, including lower maintenance efforts, lower development costs, and time to market. Unfortunately, currently implementing a framework for large-scale software reuse in Android apps is still a huge problem, regarding the complexity of the task and lacking of practical technical support from either tools or domain experts. Therefore, proposing a feature location benchmark for apps will help developers either optimize their feature location techniques or reuse the assets created in the benchmark for reusing. In this paper, we release a feature location benchmark, which can be used for those developers, who intend to compose software product lines (SPL) and release reuse in apps. The benchmark not only contributes to the research community for reuse research, but also helps participants in industry for optimizing their architecture and enhancing modularity. In addition, we also develop an Android Studio plugin named caIDE for developers to view and operate on the benchmark.
\end{abstract}

\begin{IEEEkeywords}
Software product line, feature location benchmark, reuse, android apps
\end{IEEEkeywords}

\IEEEpeerreviewmaketitle

\input{background.tex}

\input{overview.tex}

\input{feature.tex}

\input{benchconstr.tex}
\input{benchmark-organization.tex}
\input{extensions.tex}
\input{discussions.tex}

\input{conclusion.tex}

\bibliographystyle{IEEEtran}

\bibliography{reference}

\end{document}

%% file: package.tex
\usepackage{hyperref}
\usepackage{mdframed}
\usepackage{lipsum}
\usepackage{amsmath}
\usepackage{amsfonts}
\usepackage[font=small]{caption}
\usepackage{ifpdf}
\usepackage[final]{graphicx}
\usepackage{algorithmic}
\usepackage{array}

\usepackage{caption}
\usepackage{subcaption}
\usepackage{multirow}
\usepackage{float}
\usepackage{mwe}

\usepackage{url}
\newcommand{\code}[1]{{\small\texttt{#1}}}
\usepackage{todonotes}

%% file: lstingpackage.tex
\usepackage{xcolor}

\usepackage{listings}

\definecolor{light-gray}{gray}{0.80}
\definecolor{light-yellow}{rgb}{1, 0.99, 0.67}
\definecolor{light-greenyellow}{rgb}{0.85, 0.99, 0.67}

\definecolor{pink}  {rgb}{0.67, 0.05, 0.57} 
\definecolor{red}   {rgb}{0.87, 0.20, 0.00} 
\definecolor{green} {rgb}{0.00, 0.47, 0.00} 
\definecolor{violet}{rgb}{0.41, 0.12, 0.61} 
\definecolor{brown} {rgb}{0.39, 0.22, 0.13} 

\lstdefinelanguage{Objective-C 2.0}[Objective]{C} {
    morekeywords={id, Class, SEL, IMP, BOOL, nil, Nil, NO, YES,
                  oneway, in, out, inout, bycopy, byref,
                  self, super, _cmd,
                  @required, @optional,
                  @try, @throw, @catch, @finally,
                  @synchronized, @property, @synthesize, @dynamic},
    moredelim=[s][color{red}]{@"}{"},
    moredelim=[s][color{red}]{<}{>}
}

\lstdefinestyle{Xcode} {
    language        = Objective-C 2.0,
    basicstyle      = footnotesizettfamily,
    identifierstyle = color{black},
    commentstyle    = color{green},
    keywordstyle    = color{pink},
    stringstyle     = color{red},
    directivestyle  = color{brown},
    extendedchars   = true,
    tabsize         = 4,
    showspaces      = false,
    showstringspaces = false,
    breakautoindent = true,
    flexiblecolumns = true,
    keepspaces      = true,
    stepnumber      = 0,
    xleftmargin     = 0pt
}

\lstdefinestyle{C}{
	commentstyle=\color{green},
	stringstyle=\color{red},
	keywordstyle=\color{blue},
	basicstyle=\footnotesize\ttfamily,
	numbers=left,
	numberstyle=\tiny,
	numbersep=5pt,
	frame=lines,
	breaklines=true,
	prebreak=\raisebox{0ex}[0ex][0ex]{\ensuremath{\hookleftarrow}},
	showstringspaces=false,
	upquote=true,
	tabsize=2,
}

\lstdefinestyle{Java}{
	commentstyle=\color{green},
	stringstyle=\color{blue},
	keywordstyle=\color{pink},
	basicstyle=\footnotesize\ttfamily,
	numbers=left,
	numberstyle=\tiny,
	numbersep=5pt,
	frame=lines,
	breaklines=true,
	prebreak=\raisebox{0ex}[0ex][0ex]{\ensuremath{\hookleftarrow}},
	showstringspaces=false,
	upquote=true,
	tabsize=2,
}

%% file: background.tex
\section{Background}
Mobile apps dominate our daily life nowadays. Users can easily download the apps from online app stores, such as, Google Play, Amazon, Apple App Store and Windows Phone marketplace. According to a recent report from AppBrain \footnote{AppBrain:\url{www.appbrain.com}}, there are over 2,850,172 apps publicly available on Google Play until the fall of 2018. The amount of apps increased more than fourfold between 2013 and 2018. 

The huge market with a large number of active users attacks developers to contribute and release their artifacts via app stores. According to the study conducted on 4,323 apps, software reuse is the common practice among app developers. For all apps concerned, 61\% of the \code{classes} appears in two on more apps on average. Furthermore, currently reuse practice in apps mainly focuses on API library reuse, class reuse and inheritance \cite{Mojica:2014,Martin:2017}.  Unfortunately, the current reuse practice and tools cannot help developers in implementing a systematically reuse of software artifacts in terms of app development. Therefore, the current practice cannot help developer reuse software artifacts systematically. Such as, implementing a software product line (SPL) in Android apps. A software product line is a set of software-intensive systems that share a common set of features. Each system in the product line contains unique segments that are defined to fulfill specific needs of a particular market \cite{Pohl:2005}. Then, developers will create the product variants from the product line and tailor the variant to end users.

By reusing the component systematically, stakeholders can benefit from the software product line, including reducing development efforts, reducing maintenance efforts, and time to market \cite{Pohl:2005}. Unfortunately, based on some existing research experience \cite{Mojica:2014,Martin:2017,Abdalkareem:2017,Li:2017}, existing reuse practice in apps does not follow a framework with a theoretical foundation. Instead, developers are more likely to follow a naive ``copy-paste'' strategy, which may misuse some functions and even introduce potential security problems. 

To contribute the reuse practices in software product line and further help stakeholder reuse artifacts in apps, in this work, we take the first step to the systematical reuse. That is, we present an Android Studio plugin for developers to conduct the feature location, and release a benchmark of feature location. A feature in an SPL represents an user-visible function, aspect or characteristic of a system \cite{Pohl:2005}. Feature location is an important and fundamental procedure in software reuse. The task of feature location is trying to find mapping between code fragments and features \cite{Poshyvanyk:2007}. With the feature location benchmark, more research prototypes can be proposed and tested, especially the reuse framework for apps. This can help developers get arid of immature reusing practices. Also, due to the different context from desktop applications, developers, who intend to focus on proposing reuse framework for apps, should mainly reference the feature location benchmark that targets on apps.

\textbf{Contribution.} To help developers boost their reuse practice and provide back-end supports for reusing frameworks, we provide a benchmark that contains over five hundred open sourced android apps on F-Droid\footnote{F-Droid: \url{https://f-droid.org}}. Specifically, in this work, we mainly make the following contributions.
\begin{itemize}
\item We release a feature location benchmark for Android apps. The benchmark contains over five hundred apps from F-Droid.
\item We propose an Android Studio Plugin named caIDE for the benchmark. Developers can explore and even edit the benchmark with the tool provided. Moreover, developers can also use our tool to build their own benchmark and release it to the community.
\item We also discuss the uniqueness issues in feature location for apps and differences between apps and desktop applications in terms of software reuse.
\end{itemize}

\framebox{
\parbox[t][1.5cm]{0.45\textwidth}{
The benchmark of 198 Android apps with ground-truth, the scenarios, and caIDE tool can be found at:\\
\centerline{
\textbf{\url{https://sites.google.com/view/caide}},
}
including technical tutorial on how to use it.
}
}\\

\textbf{Skeleton} This paper is organized as follows: Section \ref{sec:overview} presents the overview of benchmark construction and the caIDE tool. Section \ref{sec:feature} and \ref{sec:feature_location} introduce the technical details on how to extract features, build feature models and locate features, respectively. Furthermore, Section \ref{sec:benchmark_detail} introduce the details of the benchmark in terms of features, feature models, and feature annotations. In addition, our tool provides several additional functions to allow developers further operate on the benchmark. This part is covered in Section \ref{sec:extension}. We conclude this work in Section \ref{sec:conclusion}.

%% file: overview.tex
\section{Benchmark Construction and \textit{caIDE}}\label{sec:overview}
To realize a systematic reuse in a system, the software reuse community normally propose \textit{bottom-up} techniques. Typically, in a \textit{bottom-up} approach, there are three major objectives: feature identification (Section \ref{subsec:feature_identification} and \ref{subsec:build_feature_model}), feature location (Section \ref{subsec:annotation_scheme} and \ref{subsec:feature_location}), and re-engineering \cite{Martinez:2015,Martinez:2015b}. Specifically, feature identification aims at finding and discovering features for the product line. Then, in the feature location, the feature and its implementations are mapped. For example, in a music app (e.g. JOOX music), \textit{play a music} could be a feature. The feature location approaches will find the code segments in the code base that corresponds to this feature. Finally, feature re-engineering is a transformation process that transforms the annotated system into target products. As for our benchmark, we only cover the \textit{feature identification} and \textit{feature location} process. As for the feature re-engineering, it is totally a customized procedure. That is, stakeholders and domain experts have to design the configuration for target products. However, caIDE also provide additional functions to help stakeholders on this procedure automatically. 

In general, caIDE provides following major usages: (1) annotating features in the code base; (2) viewing the annotations; and (3) other supporting functions (Section \ref{sec:extension}). All three objects in a typical \textit{bottom-up} process are covered in caIDE. Specifically, \textit{feature identification} and \textit{feature location} are covered in the annotation process. The \textit{re-engineering} process is described in supporting functions in caIDE (Section \ref{sec:extension}).

\noindent\textbf{Annotation Process}. The caIDE provides a series of functions for developers and domain experts to define features, feature model and annotate the features with visual supports. To annotate a product, caIDE guides developers through the following steps.

\begin{itemize}
\item \textbf{STEP 1.} Developers have to define the features and the feature model in the product line. (Section \ref{sec:feature})
\item \textbf{STEP 2.} Developers have to assign each feature a unique background color. Here, caIDE can automatically assign each feature a color. Developers can still customize the background color.
\item \textbf{STEP 3.} Developers can set code fragments with  features. (Section \ref{sec:feature_location})
\end{itemize}

For step 1, caIDE provides a visual support for developers to draw the feature model and edit the features in the feature model easily. After this step, a feature model file named \code{featuremodel.afm} is built. The feature model file strictly follows the feature model grammar defined by Don Batory \cite{Don:2005}, which is a well-adopted feature model representation. Then, caIDE will guide developers to assign unique color to each feature. Later, developers can assign the code fragments to features as defined in step 3. Specifically, the developers first select a code range in the editor in Android Studio IDE, then they can set the feature for the code range from the context menu. caIDE will visit the AST of the file and get the AST nodes within the code range. Then, all these nodes will be assigned to the features. Note that, if there is a ``parent-child'' relation between two AST nodes in the range, we only annotate the parent AST node. Once we remark a parent AST node to a feature, all its children are annotated the feature as well. The annotation will also be stored to external files (named $<$file$>$.color) for reusing and displaying the background color. For example, when we annotate a code segment in file \code{Reader.java}, then the caIDE will create a new file named \code{Reader.color} to store the annotation information for the file \code{Reader.java}. By putting those annotation files into the same directory with its source code, caIDE avoids the naming issue. We will introduce the corresponding technical details in Section \ref{sec:feature} and \ref{sec:feature_location}.

\noindent\textbf{Explore Process.} caIDE helps stakeholders check the annotated systems. To display the benchmark, caIDE helps developers collect all information from the annotation files(\code{.color}) in the program and renders the code fragments with background colors. More specific, caIDE will first search the \code{featuremodel.afm} to recover the feature model. Then, it will check the \code{color.json} file to build the mapping between features and colors associated. At last, caIDE will inspect all source code files and all annotation files to establish the mapping between AST nodes and features. When a user opens a file in the editor, the corresponding AST nodes will be remarked with background colors.

Far more than these fundamental solutions, we extend caIDE by providing more possible actions for developers to directly use the benchmark we provided with ease. The details of these functions are introduced in Section \ref{sec:extension}.

%% file: feature.tex
\section{Feature and Feature Model}\label{sec:feature}

\subsection{Feature and Feature Identification}\label{subsec:feature_identification}
Features in an app describe the main functions or services provided by the app, which are normally visible for end-users. As defined previously, feature names of an app should be defined by the domain experts. Normally, a domain expert has to define the features and design the feature model. However, this procedure could be time-consuming and tedious. We provide an automatic approach to recommend some features for developers. We extract the features from: (a) app's description on Google Play Store; (b) app's description for its open source repository; and (c) textual information from app's implementation. Specifically, for (c), we extract the identifiers from the code based to explore the potential feature names in the app.

\begin{figure}[htpb]
	\centering
	\includegraphics[width = 0.5\textwidth]{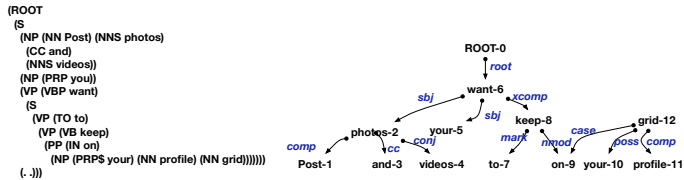}
	\caption{Parse tree of the sentence and universal dependencies between words in the sentence}
	\label{fig:gramma_tree}
\end{figure}

We use parts of speech (POS) tagging and parsing, to explore the architectures of sentences. POS tags are assigned to a single word according to its role in the sentence. Commonly-used POS tags include ADJ (i.e., adjective), VB (i.e., verb), NN (i.e., noun). Given the input sentence, we use the Stanford Parser to get the parse tree of the input sentence and the universal dependencies between words. The parse tree contains the verb phrases (i.e., VP) and noun phrases (i.e., NP) of the sentences and POS tags of each word. The universal dependency describes the relationship between words. For example, sbj refers to subject. We explore the parse tree and universal dependencies between words to identify the noun phrases and verb phrases contained in the sentence. These phrases are considered as possible features.

Specifically, for (c), we treat each \texttt{class} as flat text and use tf-idf to compute the importance of terms. The terms are ordered to be recommended as feature names. At last, feature names are selected from these recommendations.

\subsection{Building Feature Model}\label{subsec:build_feature_model}
With the features and feature annotations collected from the previous step, we build the feature model. Technically, the feature model should be built by domain experts of the systems. In caIDE, the feature model can be built manually. However, to reduce the bias and provide a handful approach for the case  that the domain expert is not available, basically we follow She et al.'s work to build the feature model \cite{She:2011} in the benchmark. She et al.'s approach \cite{She:2011} requires two inputs: the complete dependencies and extensive descriptions. The way to generate these two inputs is described as follows.
\begin{itemize}
	\item \textbf{Dependencies}: the dependencies are extracted from two aspects: (1) we manually learn the app's description and user manual (if any) and extract the potential dependencies; (2) we install the app on the emulator and try each feature in the app. Then, we describe the relations between features.
    \item \textbf{Descriptions}: the descriptions of features are collected from two parts: (1) the description, wiki and user manual of the app; (2) some descriptions of features are presented in the project's change log and commit message. When developers use some version control tools (e.g. svn, github) to manage their project, developers have to write a commit message to describe the changes for each commit.
\end{itemize}

With these information provided, the feature model can be built automatically with the approach proposed in \cite{She:2011}.

\begin{figure}[htpb]
\centering
	\includegraphics[scale=0.35]{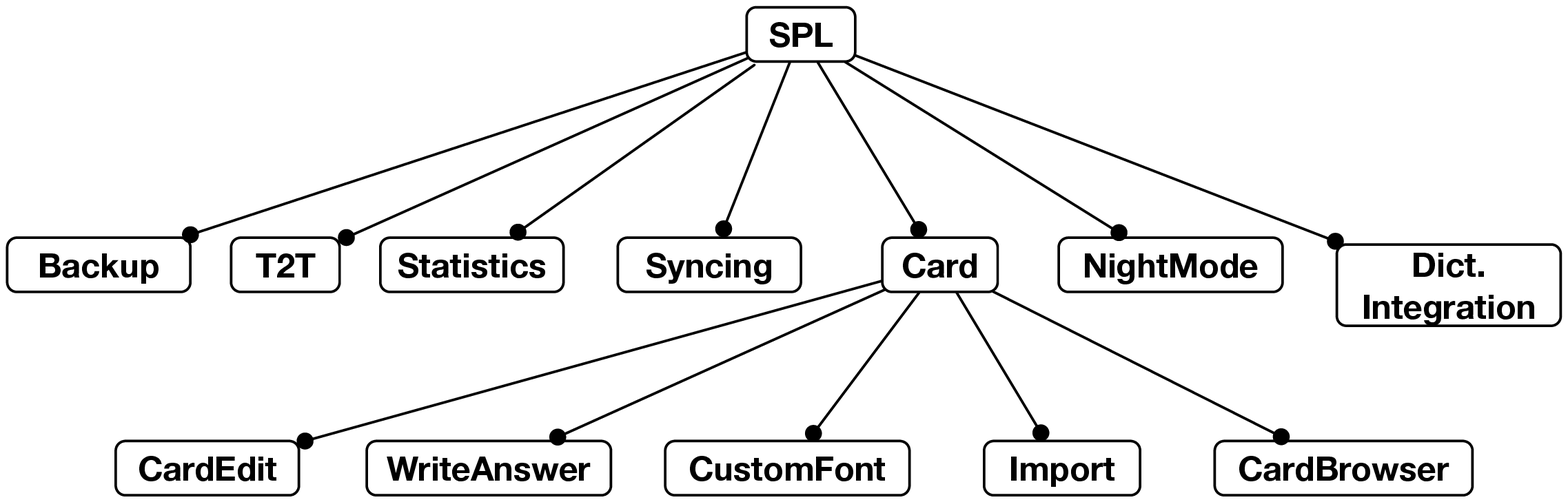}
\caption{The feature model of AnkiDroid}
\label{fig:anki_feature_model}
\end{figure}

\textbf{Example.} In the app AnkiDroid, our approach recommends 12 features, including \textit{T2T} (text-to-speach), \textit{CardBrowser}, \textit{Statistics}, \textit{NightMode}, \textit{FullBackup}, \textit{Syncing}, \textit{WriteAnswers},  \textit{WhiteBoard},  \textit{DictionaryIntegration}, \textit{CardEdit}, \textit{Import}, and \textit{CustomFont}. The feature model is built with the feature model construction approach, as shown in Figure \ref{fig:anki_feature_model}.

%% file: benchconstr.tex
\section{Feature Location}\label{sec:feature_location}

\subsection{Annotation Scheme}\label{subsec:annotation_scheme}
Prior to feature location techniques, we first introduce the annotation scheme we used in the caIDE. Our benchmark is created with our plugin named caIDE, which is built with IntellJ and targeted at Android Studio and IntellJ. With the caIDE, we can decompose the app into features, which may have a fine granularity. Developers first start with a fully composed app with all features implemented in the application. In software product line engineering, such a system is called legacy application. Then, developers can annotate code fragments with different features. One code can be associated with one or more features. To annotate code segments with features, currently there are two commonly used approaches: colored annotation \cite{Kastner:2008} and precondition compiling based annotation (a.k.a \#ifdef directive) \cite{Sincero:2010}. Specifically, the colored annotation binds code fragments with different background colors. Each color represents a unique feature defined in the feature model. 

\textbf{Example.} As shown in Listing \ref{list:colored_annotation}, different colors are associated with different features. Code fragments in the program are rendered with different colors to represent the feature annotations. Apparently, there are three features involved in the running example, include \code{push}, \code{pop}, and \code{lock}. In caIDE, we also adopt this annotation scheme.

\begin{lstlisting}[escapechar=@,language=Java,style=Java,caption={Sample of Colored Annotation},label={list:colored_annotation},captionpos=b]
class Stack{
  int size = 0;
  Object[] elementData = new Object[maxSize];
  boolean transactionsEnabled = true;
  @\colorbox{light-gray}{void push(Object o)\{}@
  	@\colorbox{light-gray}{Lock l = lock();}@
    @\colorbox{light-gray}{elementData[size++] = o;}@
    @\colorbox{light-gray}{unlock(l);}@
  @\colorbox{light-gray}{\}}@
  @\colorbox{light-yellow}{Object pop()\{}@
  	@\colorbox{light-greenyellow}{Lock l = lock();}@
    @\colorbox{light-yellow}{Object r = elementData[--size];}@
    @\colorbox{light-greenyellow}{unlock(l);}@
    @\colorbox{light-yellow}{return r;}@
  @\colorbox{light-yellow}{\}}@
  @\colorbox{light-greenyellow}{Lock lock()\{}@
  	@\colorbox{light-greenyellow}{if (!transactionsEnabled) return null;}@
    @\colorbox{light-greenyellow}{return Lock.acquire();}@
  @\colorbox{light-greenyellow}{\}}@
  @\colorbox{light-greenyellow}{void unlock(Lock lock)\{/*...*/\}}@
  @\colorbox{light-greenyellow}{String getLockVersion() \{ return "1.0";\}}@
}
@\colorbox{light-greenyellow}{class Lock \{/*...*/\}}@
\end{lstlisting}

Whereas, precondition based annotation is mainly used in product line, that developed in C/C++. In practice, developers use feature names as pre-conditions. If and only if the feature is selected, the code fragment in the directive can be covered in the target system. For example, when we use Linux system, we can configure the kernel by selecting the features we want and disabling features we are not interested. 

\textbf{Example.} In List. \ref{list:preprocessor}, there are two preconditions: \code{CONFIG\_SMP} and \code{CONFIG\_APIC}. The code \code{block 1} will be executed if the macro \code{CONFIG\_SMP} is defined. The execution condition for code \code{block 2} is \code{!CONFIG\_SMP \&\& CONFIG\_APIC}. Therefore, some benchmarks use this annotation strategy to remark the features by setting the features as the preconditions. Then, when some features are disabled in the configuration, the \code{if} condition (\code{\#ifdef CONF}) will be \code{FALSE}. The code block within the condition will not be covered in the product built.

\begin{lstlisting}[language=C,style=C,caption={Sample of Preprocessor Scheme},label={list:preprocessor},captionpos=b]
#ifdef CONFIG_SMP
//block 1
#elif defined CONFIG_APIC
//block 2
#endif
\end{lstlisting}

Currently the preprocessor annotation is also used in the benchmark in other programming languages, like Java \cite{Couto:2011}. Such annotation strategy will not affect the execution of the program, since they are always placed in the comments in the program.

In caIDE, we adopt a colored annotation strategy for our benchmark. We choose a colored annotation scheme based on two considerations.
\begin{itemize}
	\item The colored annotation will provide visual support for developers. As discussed in \cite{Feigenspan2013}, the background colors have the potential to improve program comprehension.
    \item The relations between abstract syntactic tree (AST) nodes and features are stored locally. Our tool can help developers operate the annotated system easily. Such as, inspecting the interactions between features, and displaying the code fragments for certain features.
\end{itemize}

\subsection{Feature Location}\label{subsec:feature_location}
The task of feature location is to map the features with their implementations in the code base. In our benchmark, we adopt the following steps for feature location.
\begin{itemize}
	\item \textbf{STEP 1 (recording individual feature)}. For each feature collected, we first explore the app on the physical devices and define a set of scenarios, which could represent the feature. To capture the feature at runtime, we record scenarios belonging to the feature with  \textit{Method Tracer\footnote{Method Tracer: \url{https://developer.android.com/studio/profile/am-methodtrace.html}}} in Android Debug Monitor (ADM). A set of quadruples \{name, invocation count, inclusive time, exclusive time\} in execution order is returned for representing the feature. Then, we manually annotate each features in the code base with caIDE.
    \item \textbf{STEP 2 (expand the annotations)} With the first step, only some methods for feature are annotated in the program. Then, the feature annotation process can be automatically conducted with our approach, which is a conditional probability based feature mining approach \cite{Tang:2017}. As demonstrated in the paper\cite{Tang:2017}, our feature location approach outperforms other three feature location techniques.
    \item \textbf{STEP 3 (manually checking and fixing)} Finally, we manually check the annotations returned by the feature mining approach and fix inappropriate annotations. Here, we conduct such check by reading the source code of app and running the app on emulator.
\end{itemize}

%% file: benchmark-organization.tex
\section{Benchmark}\label{sec:benchmark_detail}
In this section, we will introduce the components of the benchmark. In general, in the benchmark, we provide the following information and data for each app.

\begin{itemize}
	\item For each app, we provide a list of features and a feature model. The feature model describes the relations among these features.
    \item The feature annotation ground-truth of each project is presented as a series of \code{xml} file. Each \code{xml} file corresponds to a \code{Java} file and represents the annotations in the source code. Specifically, in the \code{xml} file, the mapping between AST nodes and features is defined. The feature location techniques can be used on the benchmark to test their performance.
    \item Our tool caIDE provides visualization of the benchmark. Code fragments for different features are assigned to different background colors. Colors assigned for each feature is described in the \code{color.json} file.
\end{itemize}

\subsection{Target Apps}
The target apps are selected from the open-source community F-Droid\footnote{F-Droid: available at \url{https://f-droid.org}}. We collect 1,365 open source Android apps from F-Droid in total. Then we apply the following criteria to select the target apps:
\begin{itemize}
	\item We only preserve those apps, that have been published via Google Play, since the apps' descriptions are mainly collected from the Google Play.
	\item We exclude apps that are not developed in Java.
    \item We exclude trial apps that have less than 3 features.
    \item For those apps, that do not have sufficient descriptions, we also skip them as well.
\end{itemize}

By applying these selecting criteria, 198 apps are reserved. These apps cover 22 categories from Google Play. The top categories include \textit{tools}, \textit{productivity}, \textit{communication} and \textit{music \& audio}. These apps' downloads ranges from 5,000 to 100,000,000. Therefore, the apps selected for our benchmark are objective.

\subsection{Features and Feature Model}
\begin{figure}[htpb]
\centering
	\includegraphics[scale=0.45]{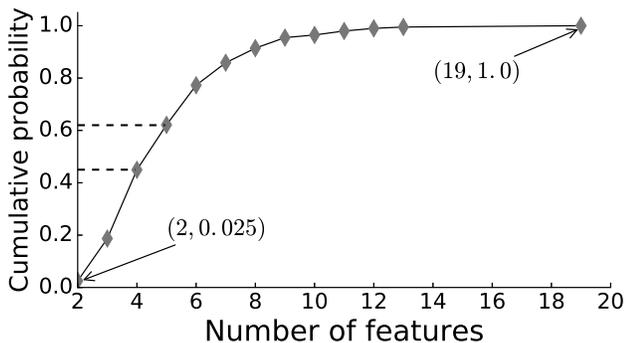}
\caption{The CDF of number of features in apps}
\label{fig:cdf_feature}
\end{figure}

In the benchmark, we use the approaches presented in Section \ref{sec:feature} to extract features and build the feature model. The feature model of the project is defined in the \code{feaeturemodel.afm} file in each project's root directory. In addition, the colors bind with features are stored in the \code{color.json} file. When developers use caIDE to update the feature model or assign/update colors for features, these two files are updated correspondingly. The number of features in each project ranges from 2 to 19. Furthermore, the cumulative distribution of the number of features in apps is shown in Figure \ref{fig:cdf_feature}. As we can observe from Figure \ref{fig:cdf_feature}, half of subjective apps have less than 6 features. This is due to some features are implemented with third-party API and some are non-functional features. Therefore, we cannot find the corresponding code fragments in the code base for these features.

By inspecting the features found from the description of the app, not all features could be considered as features for building the APL. For example, ``Fast access to items'' is a feature listed in \textit{XBMC Remote} app, but it cannot be considered as a feature, since developers do not implement an algorithm for fast accessing. In addition, those non-functional features, e.g. ``no ads ever'', ``easy to use'', also cannot be mapped to code fragments. Some features' corresponding code segments are implemented with third-party APIs. For example, in a calendar app, the description contains a sentence ``it can connect your Google Accounts and can be synchronized to automatically ...''. In practice, developers simply use the Google API to implement this function. Here, we list several features, which are frequently implemented by third-party APIs.
\begin{itemize}
	\item \textbf{Http Utility}: a lots of apps use the resources from website. Therefore, developers often have to check the connection status with http related libs. For example, the APIs \code{com.squareup.okhttp3} can be used to check the network, send the request, and diagnose.
	\item \textbf{Image download}: some apps provide functions to allow users download image from server. For example, the APIs in \code{com.squareup.picasso:picasso} provide service for image downloading and caching.
    \item \textbf{Material design}: some apps make the UI elements have a material look-and-feel design. For example, the APIs in \code{com.afollestad.material-dialogs} help developers in UI design.
    \item \textbf{Color picker}: some photo/image editing apps contains a module to let users directly select color from the paint. For example the lib \code{j4velin.colorpicker} provides such functions for users.
\end{itemize}

\subsection{Granularity}
The granularity of features in Android product lines ranges from fine to coarse. For fine granularity, a feature could be simply implemented by a \textit{field}, a \textit{statement}, or even a \textit{case} block under \textit{switch-case} structure. Whereas, some features are represented by \textit{class}es, or even \textit{package}s. Discarding those features, that are implemented with third-party APIs, we compute the proportions of granularity of other features used in apps. We find that the proportions of features in field/variable, statement, method and class are 24.6\%, 16.5\%, 60\%, 53.2\%, respectively.

\subsection{Usage of Benchmark}
The benchmark can be used by developers and researchers mainly for following cases.
\begin{itemize}
	\item \textbf{Testing feature location technique proposed.} When a novel feature location technique is proposed, there is always a need for a benchmark in order to assess the performance of the feature location technique. caIDE allows developers to design their own feature location approaches and annotate the product line with customized approaches automatically.
    \item \textbf{Building the product variants.} caIDE allows developers to build the variant applications from the annotated product line automatically. Developers only have to set the configuration from the GUI configuration panel in caIDE. Then caIDE can extract the product variants from the annotated product with the configuration provided. We introduce the details of this function in Section \ref{subsec:transform_to_variants}. 
    \item \textbf{Inspecting the feature implementation.} In addition, the benchmark can be used by developers to learn how to implement certain functions in the Android apps. For researchers, it can be used for software reuse study, programming practices, modularity, and other possible research issues.
\end{itemize}

%% file: extensions.tex
\section{Supporting Functions}\label{sec:extension}
In this section, we will introduce several additional operations already implemented in caIDE to help developers use the benchmark easily. Besides two basic modules (annotation and view), caIDE also allow developers to analyze the interactions between features (Section \ref{subsec:interaction_feature}), and extract product variants from the annotated product line.

\subsection{Interactions Between Features}\label{subsec:interaction_feature}
caIDE can also help developers analyze the interactions between features. Typically, relations and constraints between features should be described in the feature model. However, our caIDE can also help developers check the interactions and guide developers to update the feature model. For instance, if caIDE find that the execution of feature $f_{1}$ may imply the execution of feature $f_{2}$ from the annotations, it will suggest developers to add such a constraint in the feature model. Therefore, the \textit{interaction} module in our tool not only helps developers view the interactions between features, but also tries to provide suggestions for adding additional relations in the feature model. Still, the suggestions provided by the \textit{interaction} module for fixing the feature model are not compulsory and cannot be fully correct. This is because that the suggestions are merely collected from the code base and based on the program analysis, whereas the relations between features should be designed and confirmed only by the domain expert. The main purpose is to provide suggestions for developers with limited domain knowledge.

Moreover, we will introduce all interactions currently supported in the caIDE.
\begin{itemize}
	\item \textit{requires:} in an SPL, if a feature uses data from another feature, it builds a usage dependency from the data consumer to its producer. In practice, caIDE will first collect the implementations of each feature. Then, it will check whether there exists a data flow from a feature to another by traversing the program dependency graph (PDG) of the app. The PDG of an app contains both data flow dependency and control flow dependency \cite{Ferrante:1987}. 
    \item \textit{mutual exclude:} the \textit{mutual exclude} relation between two features represents that two features cannot be in the same product variant. For example, certain apps can be either in off-line mode or on-line mode. Therefore, if an app in run in an off-line mode, it always cannot in the online mode at the same time. In practice, caIDE checks such relation by detecting whether the executions of two features from two aspects: (1) whether two features are booted by different conditions; (2) whether two features are exclusive in nature from the programming perspective. For example, feature \code{f\_CHN} is represented by a class \code{Chinese} and class \code{English} represents feature \code{f\_GBR}. In addition, class \code{Chinese} and class \code{English} are inherited from class \code{Language}. Therefore, it is possible that only one feature between \code{f\_CHN} and \code{f\_GBR} can be existed in the variant product.  
\end{itemize}

Again, all these interactions are collected based on the annotations of features. The relations and constraints for features can only be used as auxiliary information for domain experts to refine the feature model. Domain experts have to confirm these recommendations and make final decisions on their own. However, this does not mean that caIDE make the incorrect conclusions. Since caIDE makes the decision only reply on the annotations for features. The relations and constraints between features are more complex and require domain knowledge. However, caIDE does not have such domain knowledge, caIDE tries to predict the relations between features based on annotations. Therefore, caIDE can only provide hints for domain experts based on annotations.

\subsection{Transforming the Annotation SPL to Variants}\label{subsec:transform_to_variants}
In addition, our caIDE can also help developers build the variants from the annotated systems (benchmark). In SPL, a variant represents a running system that can be used by end-users directly \cite{Pohl:2005}. It is built by configuring the SPL.

Recall the annotation scheme mentioned in Section\ref{subsec:annotation_scheme}, caIDE can help developers generate the target variants from the annotated system. caIDE adopts an AST-write strategy to build the variant. In general, it contains three steps. First, the code is parsed into an AST. Second, the AST nodes in the AST are assigned to different features as presented in the feature annotation. Then, based on the feature module of the app and the configuration given by the user, all AST nodes, that are associated with unwanted features in the configuration, are marked for removing from the AST. 

%% file: discussions.tex
\section{Discussions}
Furthermore, we intend to share the experience gained and point out several handful lessons learned in this work.

\textbf{Differences between desktop applications and mobile apps in terms of reuse.} We compare the annotation process in mobile apps and desktop applications, we highlight the following main differences between two types of applications.
\begin{itemize}
	\item \textbf{Call Graph vs. Window Transaction Graph.} In most desktop applications, the program starts from the \code{main} method, and then goes though the call graph based on the input context. The call graph describes the possible execution path of the program in a graph \cite{Grove:1997}. Whereas, in the mobile context, the window transition graph (WTG) is frequently used to analyze the execution of the app \cite{Yang:2015}. In the WTG, nodes represent windows and transitions between windows are linked with edges. Transitions are triggered by callbacks executed in the UI thread. Therefore, to reuse components in apps, developers not only have to take care of the execution logic, but also carefully resolve all UI elements and events involved.
    \item \textbf{Resources.} Resources (UI elements, strings and layout) are another major concern in app reuse comparing to desktop applications. Specifically, not all desktop applications have graphical user interfaces (GUI) for end users. Whereas the majority of apps have GUI provided for users. Therefore, the approach for reusing apps must carefully take care of resources. This rule is also suitable for building a software product line for Android apps. 
    \item \textbf{Third-party API.} To successfully reuse software artifacts in apps, all third-party APIs involved in the apps should also be inspected. Specifically, developers have to explore the functionality of those APIs, how these APIs are used,  how these APIs are cooperated with others, and what parts of the program are affected. We highlight the impact of third-party APIs is because that sometimes those APIs can change some existing programming practices in Android. For example, library \textit{Butter Knife} (\code{com.jakewharton:butterknife}\footnote{Butter Knife: \url{http://jakewharton.github.io/butterknife/}}) define a new approach to manage UI elements. It is apparent that an approach for building SPL for other apps cannot be applied to the app with \textit{Butter Knife}. Therefore, developers have to cope with these special cases separately. Unlike desktop application, it is hard to design a universal reuse approach for all apps. 
    \item \textbf{Android API Compatibility.} Android APIs are frequently updated, which may introduce compatibility issue in reuse. For example, after API 21, Android adopts the camera2 API to use camera resource. The camera2 API follows a different pattern comparing to camera1 in terms of using camera.
\end{itemize}

Hence, researchers, who intend to propose approaches for building SPL on Android apps, have to take into account the differences between mobile apps and desktop applications. That also means that we cannot directly use the existing approaches in software reuse and software product line within the mobile context.

\textbf{Extract product variants from annotation product line.} In addition, we can use caIDE to extract product variants from the annotation product line (benchmark). As mentioned in Section \ref{subsec:transform_to_variants}, caIDE also provides such functions to allow developers to extract the product variants from the annotation product line automatically. By conducting such experiments, we found that even caIDE can create the variant automatically, some additional efforts are still required from developers. The additional task is mainly about fixing the UI. For example, if we remove an \code{activity} in a variant based on the configuration, the button, that triggers this \code{activity}, will become invalid. Therefore, developers have to fix the code fragments related to the button. In caIDE, we leave this part for developers based on two concerns: (1) developers may want to redesign the UI and layout; and (2) rather than fixing the UI, developers can also redesign the execution logic for the variant product and keep the UI unchanged.

%% file: conclusion.tex
\section{Conclusion}\label{sec:conclusion}
In this paper, we presented caIDE, an Android Studio plugin, and a benchmark for supporting reuse software artefacts in app on feature location. The benchmark is based on several existing android apps and is designed to support research and programming practice on software reuse in the context of software product line. The benchmark can help researchers conduct reuse studies on apps rather than desktop applications. With caIDE plugin, users can directly operate on the benchmark to build the variant product or even release their own benchmarks with our tool. In the future, we intend to extend the caIDE to support more programming languages used in app development, including Kolin, python, and even native C++. In addition, we will release more benchmark for other programming languages.